\documentclass[12pt,preprint]{aastex}
\usepackage{emulateapj5}

\newcommand{\va}{V_{A}}
\newcommand{\wplus}{w^{+}}
\newcommand{\wminus}{w^{-}}
\newcommand{\chiplus}{\chi^{+}}
\newcommand{\chiminus}{\chi^{-}}

\newcommand{\wpluslam}{w^{+}_\lambda}
\newcommand{\wminuslam}{w^{-}_\lambda}
\newcommand{\chipluslam}{\chi^{+}_\lambda}
\newcommand{\chiminuslam}{\chi^{-}_\lambda}
\newcommand{\taupluslam}{\tau^{+}_\lambda}
\newcommand{\tauminuslam}{\tau^{-}_\lambda}

\newcommand{\delt}{\partial_t}
\newcommand{\bnabla}{\mbox{\boldmath $\nabla$}}

\shorttitle{Imbalanced Strong MHD Turbulence}
\shortauthors{}

\begin{document}

\title{Imbalanced Strong MHD Turbulence}
\author{Y. Lithwick$^1$, P. Goldreich$^2$, and S. Sridhar$^3$}
\affil{$^1$CITA, Toronto ON, Canada, $^2$IAS, Princeton NJ, $^3$Raman 
Research Inst., Sadashivanagar, Bangalore, India}
\email{yoram@cita.utoronto.ca, pmg@ias.edu, ssridhar@rri.res.in}

\begin{abstract} 
We present a phenomenological model of imbalanced MHD turbulence in an
incompressible magnetofluid. The steady--state cascades, of waves
traveling in opposite directions along the mean magnetic field, carry
unequal energy fluxes to small length scales, where they decay due to
viscous and resistive dissipation. The inertial--range scalings are
well--understood when both cascades are weak. We study the case when
both cascades are, in a sense, {\it strong}. The inertial--range of
this imbalanced cascade has the following properties: (i) the ratio of
the r.m.s. Elsasser amplitudes is independent of scale, and is equal
to the ratio of the corresponding energy fluxes; (ii) in common with
the balanced strong cascade, the energy spectra of both Elsasser
waves are of the anisotropic Kolmogorov form, with their parallel
correlation lengths equal to each other on all scales, and
proportional to the two--thirds power of the transverse correlation
length; (iii) 
the equality of cascade time and waveperiod (critical balance) that
characterizes the strong balanced cascade does not apply to the Elsasser field
with the larger amplitude.
 Instead, the more general
criterion that always applies to both Elsasser fields is that
the cascade time is equal to the correlation time 
of the straining imposed by oppositely-directed waves.
 (iv) in the limit that
the energy fluxes are equal, the turbulence corresponds to the
 balanced strong cascade. Our results are particularly
relevant for turbulence in the solar wind. 
Spacecraft measurements have
 established that, in the inertial range of solar wind turbulence,
  waves travelling away from the
sun have higher amplitudes than those travelling towards it.
Result (i) allows us to infer
the turbulent flux ratios from the amplitude ratios, thus providing insight into the
origin of the turbulence.

\end{abstract}

\section{Introduction}

Magnetohydrodynamic (MHD) turbulence is present in many astronomical
settings, such as the solar wind, the interstellar medium, molecular
clouds, accretion disks, and clusters of galaxies
\citep{bis03,kul05,sch05}.  Its theory has attracted a sizable
literature \citep{iro63,kra65,she83,gol95,gol97,ng96,cho00,bis00,
mar01,cho02,gal00,gal02,mul03,gal05,bol05,mul05,ber05}.  The simplest
of cases concerns the small--scale dynamics of the excitations of an
incompressible fluid with a mean magnetic field. The turbulent cascade
of energy to small scales is the result of non linear interactions
between Alfv\'en waves traveling in opposite directions along the
local, mean magnetic field \citep{iro63,kra65}. Whereas this broad
picture of Iroshnikov and Kraichnan still endures, our appreciation of
MHD turbulence has undergone significant revisions due, mainly, to the
recognition of the importance of anisotropy and the consequent
strengthening of non linear interactions. The inertial--range, which
includes length scales between the stirring and dissipation scales, is
best understood in those cases in which the oppositely directed waves
are excited with equal power: these {\it balanced} cascades can be
{\it weak} \citep{ng96,gol97}, as well as {\it strong}
\citep{gol95}. {\it Imbalanced} cascades are understood only in the
case when the turbulence is {\it weak} \citep{gal00,lit03}. In this
Letter we consider the general case of imbalanced cascades.

The solar wind is the best laboratory that we have to investigate MHD
turbulence.  
In-situ measurements by
spacecraft yield spectra for velocity and magnetic field fluctuations
over many decades of lengthscale (e.g., Horbury 1999). 
On the largest scales, the spectrum is flat, presumably reflecting the spectrum with 
which fluctuations are injected into the solar wind by shocks or dynamical instabilities.
On smaller scales, the spectrum is Kolmogorov, and fluctuations
are thought to be undergoing an active turbulent cascade.  
On these scales,
 the amplitudes of the two 
Elsasser fields are not equal:  waves travelling away from the Sun
have higher amplitudes than those travelling towards it.\footnote{This imbalance 
is more pronounced closer to the Sun.}
Because of this imbalance, the theory 
of MHD turbulence has been
inadequate for application to the solar wind.
 Our solution for the strong imbalanced cascade removes
this inadequacy. 

In \S \ref{sec:balanced}, we summarize the properties of MHD cascades
that were previously understood. Our solution for the strong
imbalanced cascade is given in \S \ref{sec:imbalanced}.

\section{Balanced cascades and the imbalanced weak cascade}
\label{sec:balanced}

The system we consider is an incompressible magnetofluid of mass density
$\rho$ and mean magnetic field $B_0\hat{z}$. Let ${\bf v}({\bf r}, t)$ be
the fluid velocity, and ${\bf b}({\bf r}, t)$ the magnetic field
fluctuation. The MHD equations determining their time evolution are
expressed conveniently in terms of the Elsasser fields, ${\bf w}^\pm
=
{\bf v}\mp{\bf b}/\sqrt{4\pi\rho}:\ $\footnote{
The superscripts on ${\bf w}^+$, ${\bf w}^-$ refer to the direction of wave propagation.
}
\begin{eqnarray}
(\delt \pm\va\partial_z){\bf w}^\pm
\;+\; ({\bf w}^\mp\cdot\bnabla)\,{\bf w}^\pm &\;=\;&
-\bnabla\left(p/\rho\right) \;
\label{eq:mhd} 
\end{eqnarray}
\noindent where $\va = B_0/\sqrt{4\pi\rho}$ is the Alfv\'en speed, and $p$
is the total pressure, determined by requiring $\bnabla\cdot{\bf
w}^{\pm} = 0\,$.
 We have neglected
viscous and resistive dissipative terms, which are important on
small scales; we have also neglected forcing terms, which
are important on large scales. 
 When either ${\bf w}^+$ or
${\bf w}^-$ is initially zero everywhere, the nonlinear term
vanishes for all time. Then, either $\left\{{\bf w}^- = {\bf
w}^-(x, y, z + \va t),\;{\bf w}^+ = {\bf 0}\right\}$,
or $\left\{{\bf w}^+ = {\bf w}^+(x, y, z - \va t),\;{\bf
w}^- = {\bf 0}\right\}$, is a nonlinear solution of arbitrary
form, that propagates in the $-\hat{z}$ or $+\hat{z}$ direction with
speed $\va$. Kraichnan recognised that the existence of these
nonlinear solutions implies that MHD turbulence can be
described as interactions between oppositely directed
wavepackets. Equations~(\ref{eq:mhd}) conserve the Elsasser
energies, $E^{\pm} =\frac{1}{2} \int\, \left|{\bf
w}^{\pm}\right|^2\,d^3x\,$, hence collisions between
wavepackets do not lead to exchanges between $E^+$ and
$E^-$, but only to a redistribution of the energies over
different length scales.

We label the r.m.s. amplitudes
of the Elsasser fields in the inertial range of the turbulence by $w^\pm_\lambda$, where $\lambda$ is the lengthscale
transverse to the mean B-field.
MHD turbulence is known to be anisotropic: wavepackets of each type of Elsasser field
with transverse scale $\lambda$ have an  
extent $\Lambda_\lambda^\pm$
parallel to the mean B-field.\footnote{A wavepacket  can be defined more precisely
in terms of correlation lengths
perpendicular and parallel  to the mean field \citep{lit01}.}
We note that  $\Lambda_\lambda^\pm$ is scale-dependent, and 
is typically much larger than $\lambda$.
We define the
 strength of a wavepacket as the ratio of the straining rate that it imposes
 on oppositely directed waves,
$w^{\pm}_\lambda/\lambda\,$,
to its linear wave frequency:
\begin{equation}
\chi^{\pm}_\lambda \;\equiv\; \frac{\Lambda^{\pm}_\lambda
w^{\pm}_\lambda}{\lambda\va}\, .
\label{chidef}
\end{equation}
If a wavepacket of $+$ve waves has  $\chi_\lambda^+=1$, 
then $-$ve waves with scale $\lambda$ that encounter it are cascaded in 
the time that they cross it, assuming that the backreaction of the $-$ve waves
on the $+$ve ones is negligible.We wish to understand how $\chi^{\pm}_\lambda$
influence the
inertial--range scalings of physical quantities. Below we provide a brief
review of the cases that are understood to some extent; this enables us to
pose our question more sharply.\footnote{In the balanced cases, we set $\wplus_\lambda=
\wminus_\lambda=w_\lambda\,$ and $\Lambda_\lambda^+=\Lambda_\lambda^-=\Lambda_\lambda$ 
which implies that $\chiplus_\lambda= \chiminus_\lambda = \chi_\lambda=
(\Lambda_\lambda w_\lambda/\lambda\va)$.} 

\noindent
1. {\it Balanced weak turbulence}, $\chi_\lambda\ll 1\,$: The turbulent cascade
is due to resonant $3$--wave interactions \citep{ng96,gol97,gal00,lit03}.
The r.m.s.
amplitude across a transverse scale $\lambda$ is $w_\lambda\propto
\lambda^{1/2}$. Frequency resonance conditions prevent a parallel cascade,
so the parallel scale is independent of $\lambda$; its value is set by conditions
at the stirring scale.
  Perturbation
theory is valid when $\chi_\lambda\ll 1$, and can be used to derive
kinetic equations, describing the inertial--range. However, the cascade
strengthens on small scales, because $\chi_\lambda\propto1/\lambda^{1/2}$
increases as $\lambda$ decreases; this limits the validity of perturbation
theory, and the inertial range of weak turbulence. 

\noindent
2. {\it Balanced strong turbulence}, $\chi_\lambda\sim 1\,$: Perturbation theory
is inapplicable. According to the phenomenological theory of
\citet{gol95}, the strength of the interactions remains of order unity:
$\chi_\lambda\sim 1$ (``critical balance''). The cascade is of anisotropic
Kolmogorov form, with $w_\lambda\propto\lambda^{1/3}$ and
$\Lambda_\lambda\propto \lambda^{2/3}\,$.   Turbulence that is excited
with $\chi_\lambda>1$, weakens to $\chi_\lambda\sim 1$ in 
less than the Alfv\'en crossing time.

\noindent
3. {\it Imbalanced weak turbulence},
$\chi_\lambda^\pm\ll 1$:
 In common with the balanced case, the process can be described
in detail, using the weak turbulence theory of resonant 3--wave
interactions \citep{gal00,lit03}. As before, 
the product $w_\lambda^+w_\lambda^-\propto \lambda$, and
frequency resonance
conditions forbid a parallel cascade, so the two parallel scales
$\Lambda_\lambda^\pm$ are scale-independent.
However, in contrast to the balanced case, (i) kinetic
equations are required to relate the spectral indices of the
oppositely directed waves to the ratio of their fluxes; (ii) the ratio
of $\wpluslam$ to $\wminuslam$ in the inertial range depends on the
ratio of $\lambda$ to the dissipation scale, $\lambda_{\rm diss}$,
since the two amplitudes are forced to equal one another at
$\lambda_{\rm diss}$.

As discussed in \citet{lit03}, when the dissipation scale is large
enough (while still remaining smaller than the stirring scale),
imbalanced turbulence can be weak throughout the
inertial--range. However, both $\chipluslam$ and $\chiminuslam$
increase with decreasing $\lambda$, thereby limiting the inertial
range. Therefore, in the physically important limit of very small
dissipation, it is inevitable that at least one of the $+$ve or
$-$ve cascades becomes strong.
 
\section{The imbalanced strong cascade}
\label{sec:imbalanced}

In the present section, the heart of this paper, we derive the spectrum
of imbalanced strong turbulence.
We assume that $w_\lambda^+\gg w_\lambda^-$.\footnote{This assumption
is made only for clarity of presenation.  Our discussion is more
generally
valid, for all $w_\lambda^+\geq w_\lambda^-$. }

Let $\chi_\lambda^+\geq 1$.  Otherwise, the turbulence would be weak.
The $+$ve waves impose a strain on $-$ve ones, with straining rate
$\sim \wpluslam/\lambda$.
Since
$\chipluslam\geq 1$, this strain is nearly constant over the
cascade time, $\tauminuslam$, of the $-$ve waves. 
Thus
$\tauminuslam \sim \lambda/ \wpluslam$ and the energy flux in the
$-$ve cascade is
\begin{equation}
\varepsilon^{-}\;\sim\;
\frac{(w_\lambda^-)^2}{\tau_\lambda^-}\;\sim\;
\frac{(w_\lambda^-)^2w_\lambda^+}{\lambda}\, .
\label{eq:fluxminus}
\end{equation}
Waves of $w_\lambda^-$ travel a parallel distance $\va\tauminuslam$ before cascading to
smaller $\lambda$.  It {\it follows} that a wavepacket with  given transverse size $\lambda$ has
parallel size
\begin{equation} 
\Lambda^-_\lambda \;\sim\; \va\tauminuslam \;\sim\;
\left(\frac{\va}{\wpluslam}\right)\lambda\,.
\label{eq:Lambdaminus}
\end{equation}
\noindent
Since this relation mixes $\Lambda^-_\lambda$ and $w_\lambda^+$, it does not
amount to critical balance for either the $+$ve or $-$ve waves. 

Equation ({\ref{eq:Lambdaminus}}), together with our starting point $\chi_\lambda^+\geq 1$, 
imply that $\Lambda_\lambda^+\gtrsim\Lambda_\lambda^-$.
But, since $-$ve waves backreact on the $+$ve ones, they impose uncorrelated strains on parts of
a $+$ wavepacket separated in $z$ by
$\ga\Lambda^-_\lambda$. Thus they imprint their parallel
scales on the $+$ve waves, and
\begin{equation}
\Lambda^+_\lambda \;\sim\;\Lambda^-_\lambda \ .
\label{eq:equal}
\end{equation}
\noindent
Henceforth we denote the common parallel scales by
$\Lambda_\lambda\,$.  
This {\it discovery} that the parallel scales are similar in an imbalanced cascade is a 
{\it nontrivial}
result.
Equations (\ref{eq:Lambdaminus}) and
(\ref{eq:equal}) yield
\begin{equation} 
\chipluslam \;=\; \frac{\wpluslam\Lambda_\lambda}{\va\lambda}
\;\sim\; 1\, \ ,
\label{eq:critical}
\end{equation}
\noindent
{\it proving} that the cascade of $-$ve waves is critically balanced.

It remains to calculate the $+$ve waves' cascade time.
All of the following material up to equation (\ref{eq:fluxplus2}), as 
well as the material in the Appendix, is devoted to  {\it deriving} the result:
$\tau_\lambda^+\sim\lambda/w_\lambda^-$.  
This  result is remarkable.
It shows that the straining rate imposed
by the $-$ve waves on $+$ve ones, $w_\lambda^-/\lambda$, is imposed coherently
over a time $\lambda/w_\lambda^-$.  Yet the waveperiod of $-$ve waves is much shorter than this,
by the factor
\begin{equation} 
\chiminuslam \;=\; \frac{\wminuslam\Lambda_\lambda}{\va\lambda}
\;\sim\; \frac{\wminuslam}{\wpluslam} \; \ll 1\, .
\label{eq:subcritical}
\end{equation}
Since $\chi_\lambda^-\ll 1$, one might be tempted to conclude, erroneously, 
that $+$ve waves undergo a weak cascade, i.e., 
that
 $-$ve waves impose on them small,
short uncorrelated strains of amplitude $\chi_\lambda^-$
 over time intervals $\sim
\Lambda_\lambda/\va$ resulting in $\tau_\lambda^+\sim
(\chi_\lambda^-)^{-2}(\Lambda_\lambda/\va)\sim
(\chi_\lambda^-)^{-1}(\lambda/w_\lambda^-)$.  Instead, the correct conclusion
is that a coherent strain is imposed over  time interval $\lambda/w_\lambda^-$.

How can the coherence time 
exceed $\Lambda_\lambda/\va$?  The key point
is
that the straining of the $+$ve waves is due to the ${\bf
w}^-$ field as seen from the $+$ve waves' rest frame
(which has $x'=x, \;y'=y, \;z' = z - \va t, \;t'=t$). In this frame,
the MHD equations~(\ref{eq:mhd}) transform to
\begin{eqnarray}
\partial_{t'}{\bf w}^+
\;+\; ({\bf w}^-\cdot\bnabla')\,{\bf w}^+ &\;=\;&
-\bnabla'\left(p/\rho\right) \;
\nonumber\\ 
(\partial_{t'} -\, 2\va\partial_{z'}){\bf w}^-
\;+\; ({\bf w}^+\cdot\bnabla')\,{\bf w}^- &\;=\;&
-\bnabla'\left(p/\rho\right) \;
\label{eq:mhdrest} 
\end{eqnarray}
To appreciate that the correlation time of ${\bf w}^-$
in the primed frame can exceed $\Lambda_\lambda/\va$, consider the
limiting case in which ${\bf w}^-$ is so small that backreaction onto 
the $+$ve waves can be neglected.  Then
${\bf w}^+$ is independent of $t'$, and is a function only of
${\bf r}'$, so ${\bf w}^-$ satisfies a linear
(integro--differential) equation, whose 
coefficients are independent of $t'$. 
If $-$ve waves are injected on a  lengthscale much larger than 
the scale of interest, with a long coherence time $T$, 
 then as they cascade down to smaller scales
their coherence time remains fixed,
$\tau_{\rm corr,\lambda}^-=T$,
where $\tau_{\rm corr,\lambda}^-$ is defined as the correlation time of the
$-$ve waves in the frame of the $+$ve waves.  
In the limiting case that  ${\bf w^-}$ is held fixed at the injection
scale ($T=\infty$),  then on smaller scales ${\bf w}^-$ is independent of $t'$
($\tau_{\rm corr,\lambda}^-=\infty$), even though it is undergoing
an active cascade to small scales.

To estimate $\tau_{\rm corr,\lambda}^-$ when ${\bf
w}^-$ is not infinitesimally small, it is necessary to
account for backreaction: $-$ve waves alter $+$ve waves,
which react back on the $-$ve ones.  As $-$ve waves
cross a plane at fixed $z'$, the $+$ve waves at that plane are
changing on their cascade time scale $\tau_\lambda^+$.  Hence,
over times separated by $\tau_\lambda^+$, the
$-$ve waves crossing $z^\prime$ are cascaded by entirely
different $+$ve waves.  This implies that $\tau_{\rm
corr,\lambda}^-\sim \tau_\lambda^+$.\footnote{We are
implicitly assuming that most of the change in $w_\lambda^-$
is accumulated on scales comparable to $\lambda$.  We justify this
assumption quantitatively in the Appendix.}  Because the $+$ve waves
are strained at rate $w_\lambda^-/\lambda$, it
follows that $\tau_\lambda^+\sim\lambda/w_\lambda^-$.
Thus
\begin{equation}
\varepsilon^{+}\;\sim\;
\frac{(w_\lambda^+)^2}{\tau_\lambda^+}\;\sim\;
\frac{(w_\lambda^+)^2w_\lambda^-}{\lambda} \,.
\label{eq:fluxplus2}
\end{equation}
Invoking Kolmogorov's hypothesis of the scale (i.e. $\lambda$) independence
of the energy fluxes given by equations~(\ref{eq:fluxminus}) and
(\ref{eq:fluxplus2}),  we obtain the inertial--range scalings,
\begin{equation}
w^{\pm}_\lambda \;\sim\; \frac{(\varepsilon^{\pm})^{2/3}}{(\varepsilon^{\mp})^{1/3}}\lambda^{1/3} \,;\qquad \Lambda_\lambda \;\sim
\frac{(\varepsilon^-)^{1/3}}{(\varepsilon^+)^{2/3}}\va\lambda^{2/3}\,.
\label{eq:scalings}
\end{equation}

The $+$ve wave cascade shares some characteristics with both weak
and strong balanced MHD cascades.  In the weak cascades, the cascade time
is longer than the waveperiod, and a wave experiences multiple,
randomly-phased perturbations during its cascade time.  In the strong
cascades, the cascade time is comparable to (or shorter than) the waveperiod, and a wave suffers a coherent strain as it cascades.
Furthermore, weak turbulence submits to perturbation theory but strong
turbulence does not. In the $+$ve wave cascade:
\begin{itemize}
\item The period of a $+$ve wave is shorter than its
cascade time.
\item A $+$ve wave is coherently strained as it
cascades.\footnote{The correlation time of the strain induced by the
$-$ve waves, $\tau_{\rm corr,\lambda}^-$, is comparable to
the
cascade time, $\tau_\lambda^+$, of the $+$ve waves.} 
\item The $+$ve wave cascade is non-perturbative. 
\end{itemize}
We contend that the $+$ve wave cascade is strong because the second
and third items have dynamical significance whereas the first does
not. The dimensionless parameter that indicates whether the
$+$ve waves are strongly cascaded is
\begin{equation}
\hat{\chi}_\lambda^{-}\equiv {w_\lambda^-
\tau_{\rm corr,\lambda}^-
\over \lambda} \ ,
\end{equation}
and not $\chi_\lambda^-$. Strong cascades correspond to
$\hat{\chi}_\lambda^{-}\sim 1$, and weak ones to
$\hat{\chi}_\lambda^{-}<1$. 
For the $-$ve wave cascade, $\hat{\chi}_\lambda^+={\chi}_\lambda^+\sim 1$, since the
correlation time of $+$ve waves in the frame of the $-$ve ones is $\Lambda_\lambda/\va$.
We call the criterion
$\hat{\chi}_\lambda^\pm\sim 1$ ``modified critical balance,'' 
to distinguish
it from critical balance (which would incorrectly imply $\chi_\lambda^\pm\sim 1$).

\section{Summary}
\label{sec:summary}
We have deduced the behavior of imbalanced strong MHD turbulence.
Its salient properties are:

\noindent
1. The $+$ve and $-$ve waves carry unequal energy fluxes,
$\varepsilon^+\neq \varepsilon^-\,$, while they both
undergo {\it strong} cascades.

\noindent
2. In the inertial--range, the r.m.s. Elsasser amplitudes are proportional
to the one--third power of the transverse scale: $w^{\pm}_\lambda\propto
\lambda^{1/3}\,$. This is similar to the balanced, strong
cascade. Moreover, their ratio, $\wpluslam/\wminuslam \sim
\varepsilon^+/\varepsilon^-$ is independent of $\lambda$.

\noindent
3. The parallel scales of the $+$ve and $-$ve waves are
 equal. The common parallel scale of eddies of transverse scale
$\lambda$, is $\Lambda_\lambda \propto\lambda^{2/3}$, similar to the
balanced, strong cascade.

\noindent
4. The cascade times of the (larger amplitude) $+$ve waves are longer
than their waveperiods by the constant factor
$(\varepsilon^+/\varepsilon^-) \geq 1\,$, independent of
scale: unlike the imbalanced, weak cascade, there is no tendency for the
cascade to strengthen at small $\lambda$. When $\varepsilon^+
=\varepsilon^-\,$, the the turbulence corresponds to the
balanced strong cascade of \citet{gol95}.

\newpage
\appendix
\section{Backreaction and the Correlation Time of the $-$ve Waves}
Let $\tau_\lambda^+\sim\Lambda'_\lambda/\va$ denote the time scale
over which $\wpluslam$ varies at fixed $z'$. Below we prove that (i)
the correlation time of the $-$ve waves, at fixed $z'\,$, is
$\tau_{\rm corr,\lambda}^- \sim \tau_\lambda^+\,$; (ii)
$\Lambda'_\lambda/\Lambda_\lambda\sim \varepsilon^{+}/
\varepsilon^{-}\,$. 

During the time interval
$\tau_{\lambda_*}^+\sim \Lambda'_{\lambda_*}/\va$,
$-$ve waves cascade from transverse scales $\lambda_{\rm max}$
to $\lambda_*$. We approximate this cascade as taking place in
discrete steps of duration $\tau_\lambda^-\sim
\Lambda_\lambda/\va$ in each of which $\lambda$ decreases by a
constant factor of order unity. Thus $\lambda_{\rm max}$ is related to
$\lambda_*$ by
\begin{equation}
\int_{\lambda_*}^{\lambda_{\rm max}}\,\frac{d\lambda}{\lambda}\,\Lambda_\lambda
\;\sim\; \Lambda'_{\lambda_*}
\label{eq:lambdaprime}
\end{equation}
\noindent
If $\Lambda'_\lambda$ increases with $\lambda$,
equation~(\ref{eq:lambdaprime}) implies that
\begin{equation}
\Lambda'_{\lambda_*} \;\sim\; \Lambda_{\lambda_{\rm max}}\; ;
\label{eq:lambdastar}
\end{equation}
\noindent most of the cascade time is spent near scales
$\sim\lambda_{\rm max}$.

The $+$ve waves encountered by $-$ve waves, at the same
$z'$ and
$\lambda$, but separated in time by $\tau_{\lambda_*}^+$, differ
by $(\delta\wpluslam/\wpluslam)\sim
\Lambda'_{\lambda_*}/\Lambda'_\lambda$.\footnote{We reserve the symbol
$\delta$ to denote differences accrued over the time interval
$\tau_{\lambda_*}^+$.}  Thus over each step of the cascade from
$\lambda_{\max}$ to $\lambda_*$, 
\begin{equation}
\delta\wminuslam \;\sim\; \frac{\delta\wpluslam}{\lambda}
\frac{\Lambda_\lambda}{\va}\,\wminuslam \;\sim\;
\frac{\Lambda'_{\lambda_*}}{\Lambda'_\lambda}\,\wminuslam\, ,
\label{eq:deltawminus}
\end{equation}
\noindent where we have used, $\chipluslam \sim (\wpluslam
\Lambda_\lambda/ \va\lambda) \sim 1\,$.  Then the mean square
fractional variation of $w^-_{\lambda_*}$ accumulated during the
the entire cascade amounts to
\begin{equation}
\left(\frac{\delta
w^-_{\lambda_*}}{w^-_{\lambda_*}}\right)^2 \;\sim\;
\int_{\lambda_*}^{\lambda_{\rm max}}\,\frac{d\lambda}{\lambda}\,
\left(\frac{\Lambda'_{\lambda_*}}{\Lambda'_\lambda}\right)^2
\left(\frac{\wminuslam}{w^-_{\lambda_*}}\right)^2\,.
\label{eq:meansq}
\end{equation}
\noindent
Provided $\Lambda'_\lambda/\wminuslam$ increases with $\lambda$, 
\begin{equation}
\frac{\delta w^-_{\lambda_*}}{w^-_{\lambda_*}} \;\sim\;
1\,.
\label{eq:fracchange}
\end{equation}
\noindent
Although most of the cascade time is spent near $\lambda_{\rm max}$
(see equation~\ref{eq:lambdastar}), the accumulated change in
$w^-_{\lambda_*}$ comes from scales near
$\lambda_*$. Moreover, since $w^-_{\lambda_*}$ varies by
order unity, at fixed $z'$, during the time interval, $\taupluslam$,
the correlation time of the $-$ve waves is,
\begin{equation}
\tau_{\rm corr,\lambda}^- \;\sim\; \taupluslam \;\sim\;
\frac{\lambda}{\wminuslam}\,,
\label{eq:corrtime}
\end{equation}
\noindent 
which proves item (i). To prove (ii), we note that
\begin{equation}
\frac{\Lambda'_\lambda}{\Lambda_\lambda} \;\sim\; 
\frac{\taupluslam}{\tauminuslam} \;\sim\;
\frac{\wpluslam}{\wminuslam} \;\sim\;
\frac{\varepsilon^+}{\varepsilon^-} \;\geq\; 1\,,
\label{eq:ratios2}
\end{equation}
\noindent
where we have used equations~(\ref{eq:corrtime}) and (\ref{eq:scalings}).

\end{document}